\documentclass[usenatbib,usegraphicx]{mn2e}
\def\FF{{\rm FF}}

\def\mnras{MNRAS}
\def\aj{AJ}
\def\apj{ApJ}
\def\apjl{ApJL}
\def\aap{A\&A}
\def\pasj{Publications of the Astronomical Society of Japan}
\def\procspie{Proceedings of the SPIE}

\usepackage{epsfig}

\begin{document}
\title[Luminosity, colour and morphology dependence of galaxy
  filaments in SDSS DR4]
{ The Luminosity, Colour and Morphology dependence of galaxy filaments in  the Sloan Digital Sky Survey Data Release Four}  

\author[B. Pandey and S. Bharadwaj ] {Biswajit Pandey\thanks{Email: 
    pandey@cts.iitkgp.ernet.in} and Somnath Bharadwaj\thanks{Email: 
    somnathb@iitkgp.ac.in}
\\ Department of Physics and Meteorology\\ and
\\ Centre for Theoretical Studies \\ IIT Kharagpur \\ Pin: 721 302 ,
India }
\maketitle
\begin{abstract}
We have tested for luminosity, colour and morphology dependence of the
degree of filamentarity in seven nearly two dimensional
strips from the Sloan Digital Sky Survey Data Release Four (SDSS
DR4). The analysis is carried out at various  levels of coarse
graining allowing us to address different length-scales. We find that
the brighter galaxies have a less  filamentary distribution than the
fainter ones at all levels of coarse graining. The distribution of red
galaxies and ellipticals shows  a higher degree  of filamentarity 
compared to blue galaxies and spirals respectively 
at low levels of coarse graining. The behaviour is reversed at higher 
levels of coarse graining. We propose a  picture where the ellipticals
are densely distributed in the vicinity of the nodes where the
filaments intersect while the spirals are sparsely distributed along
the entire extent of the filaments. Our findings indicate that the
regions with an excess of ellipticals  are larger than galaxy
clusters, protruding  into the filaments. We have also compared 
the predictions of a semi-analytic model of galaxy formation
(the Millennium Run galaxy catalogue) against our results for the
SDSS. We find  the two to be in agreement for the $M^{*}$ galaxies and
for the red galaxies, while the model fails to correctly predict the 
filamentarity of the brighter galaxies and the blue galaxies.
\end{abstract}


\section{Introduction}

One of the most significant aim of all large redshift surveys is to
determine the spatial distribution of galaxies in the Universe. Over
the last few decades redshift surveys have revealed the large scale
structures in the Universe in it's full glory (e.g. , CfA ,
\citealt{gel}; LCRS, \citealt{shect}; 2dFGRS, \citealt{colles} and
SDSS, \citealt{stout}, \citealt{abaz1}, \citealt{abaz2}). It is found
that the vast majority of galaxies preferentially reside in an
intricate network of interconnected filaments and walls surrounded by
voids. The filaments have a dimension of many tens of {\rm Mpc} in
length (\citealt{bharad2}) and 1-2 $\,{\rm Mpc}$ in thickness
(\citealt{rat}, \citealt{doro2}). In fact the Local Group in which the
Milky Way resides, is believed to lie along an extended filament that
originates near the Ursa Major and includes both the IC342 and M81
groups and connects with another primary filament which originates
from the Virgo cluster and includes the Sculpture group
(\citealt{peeb}).  The complex filamentary network of galaxies, the so
called 'Cosmic web' is possibly the most striking visible feature in
the large scale structure of the Universe. Quantifying the Cosmic web
and tracing it's origin has remained one of the central issues in
cosmology.  The analysis of filamentary patterns in the galaxy
distribution has a long history dating back to a few papers in the
mid-eighties by \citet{zel}, \citet{shand1} and \citet{einas1}. More
recently, extensive studies of the LCRS reveal a rich network of
interconnected filaments surrounding voids (eg. \citealt{shand2} ;
\citealt{bharad1} ; \citealt{mul} ; \citealt{doro1} ; \citealt{trac} ;
\citealt{doro2} ; \citealt{bharad2}). 
\citet{bharad3} have compared the filamentarity in the LCRS galaxy
distribution with $\Lambda$CDM dark matter N-body simulations to  show
that the two are consistent provided  a mild galaxy bias is included.  
 \citet{sheth2} has used a technique SURFGEN
(\citealt{sheth1}) to study the geometry, topology and morphology of
the superclusters in mock SDSS catalogues. \citet{basil} and
\citet{koloko} have studied the super-cluster void network in the PSCz
and the Abell/ACO cluster catalogue respectively, finding
filamentarity to be the dominant feature. \citet{pimb} have studied
the intercluster galaxy filaments in the 2dFGRS.

The Sloan Digital Sky Survey (SDSS) \citep{york} is currently the
largest galaxy redshift survey. In a recent paper \citep{pandey} we
have analyzed the filamentarity in the two equatorial strips of the
First Data Release of the SDSS (SDSS DR1).  These strips are nearly
two dimensional (2-D). We have projected the data onto a plane and
analyzed the resulting 2-D galaxy distribution.  The large volume and
dense sampling of the SDSS allows us to construct volume limited
subsamples extending over lengthscales which are substantially larger
than possible with earlier surveys.  We find evidence for connectivity
and filamentarity in excess of that of a random point distribution,
indicating the existence of an interconnected network of
filaments. The filamentarity is found to be statistically significant
up to length-scales $80 \, h^{-1} {\rm Mpc}$ and not beyond
\citep{pandey}. Further, we show that the degree of filamentarity
exhibits a luminosity dependence with the brighter galaxies having a
more concentrated and less filamentary distribution.

It is now quite well accepted that galaxies with different physical
properties are differently distributed in space.  Studies over several
decades have established that ellipticals and spirals are not
distributed in the same way. The ellipticals are found predominantly
inside regular, rich clusters whereas the field galaxies are mostly
spirals. This effect is referred to as ``morphological segregation''
which has been studied extensively in the literature
(eg. \citealt{hubble}, \citealt{zwicky}, \citealt{davis2},
\citealt{dress} , \citealt{guzo}, \citealt{zevi},
\citealt{goto}). Further, ellipticals exhibit a stronger clustering as
compared to spirals (\citealt{zevi}). The analysis of the two-point
correlation function of galaxies with different colours shows the red
galaxies, which are mainly ellipticals with old stellar populations,
to have a stronger clustering as compared to the blue galaxies which
are mostly spirals (e.g. \citealt{will}, \citealt{brown},
\citealt{zehavi}). Studies of the topology using the genus statistics
show that red galaxies exhibit a shift towards a meatball topology
(\citealt{hoyle1}, SDSS EDR, \citealt{park1}, SDSS DR3) implying that
they prefer to inhabit the high density environments. It is found that
luminous galaxies exhibit a stronger clustering than their fainter
counterparts (e.g. \citealt{hamil}, \citealt{davis1}, \citealt{white},
\citealt{park}, \citealt{love}, \citealt{guzo}, \citealt{beno},
\citealt{nor}, \citealt{zehavi}). The difference increase markedly
above the characteristic magnitude $M*$ (\citealt{nor}, 2dFGRS,
\citealt{zehavi}, SDSS DR2) of the Schecter luminosity function
(\citealt{sek}). The detailed luminosity dependence has been difficult
to establish because of the limited dynamic range of even the largest
redshift surveys.  

\citet{einas2} have studied the luminosity distribution of galaxies in
high and low density regions of the SDSS to show that brighter
galaxies are preferentially distributed in the high density
environments. \citet{goto} have studied the morphology density
relation in the SDSS(EDR) and found that this relation is less
noticeable in the sparsest regions indicating the need for a denser
environment for fostering the physical mechanisms responsible for the
galaxy morphological changes. \citet{hog1} and \citet{blan1} find a
strong environment dependence for both the colour and luminosity for
the SDSS galaxies.

The dependence of clustering on galaxy properties like the luminosity,
colour and morphology provides very important inputs for theories of
galaxy formation. It is a natural prediction of hierarchical structure
formation that the rarer objects which correspond to peaks in the
density field should be more strongly clustered than the population
itself (\citealt{kais}, \citealt{davis3}, \citealt{white1}). A
possible interpretation of the observations is that the more luminous
galaxies are hosted in higher density peaks as compared to the fainter
galaxies. In this scenario the properties of a galaxy are largely decided
by the initial conditions at the location where the galaxy is
formed. This idea is implemented in the halo model (\citealt{jingmo},
\citealt{benson}, \citealt{seljak}, \citealt{pkok}, \citealt{ma},
\citealt{scocci2}, \citealt{white1}, \citealt{berlind},
\citealt{scran}, \citealt{yang1}, \citealt{yang2}) where it is
postulated that all galaxies lie in virialised halos and the number
and type of galaxies in a halo is entirely determined by the halo
mass. \citet{cooray} provide a detailed review of the halo model. It
is well known that galaxy-galaxy interactions and environmental effects
are also important in determining the physical properties of a
galaxy. For example, ram pressure stripping in galaxy clusters is a
possible mechanism for the transformation of a spiral galaxy into an
elliptical. Galaxy-galaxy interactions could also provide a mechanism
for morphological transformation.  In the classical picture galaxies
evolve in isolation and morphology is determined at birth, while in
the hierarchical model galaxies have no fixed morphology and they can
evolve depending on the nature of their mergers. A study of the
spatial distribution of different kind of galaxies holds the potential
to probe the factors responsible in determining galaxy luminosity,
colour and morphology.

Much of the research on the luminosity, morphology and colour
dependence of the galaxy distribution has focused either on the
morphology segregation in clusters or has studied the behaviour of the
two point correlation as a function of these galaxy
properties. Filaments are the largest known statistically significant
coherent features. They have been shown to
be statistically significant to scales as large as $80 \, h^{-1} \,
{\rm Mpc}$. In this paper we  study  the luminosity, colour and
morphology dependence of filaments  in the Sloan Digital Sky Survey
Data Release Four (SDSS DR4). This allows us to study the 
distribution of different types of galaxies on the largest
length-scales possible. Such a study will enable us to estimate  the
length-scale over which the factors responsible for the galaxy
segregation operate. While the two-point correlation function
completely characterizes the statistical properties of a Gaussian
random field, it is well known that the galaxy distribution is
significantly non-Gaussian. In fact the presence of large-scale
coherent features like filaments is a clear indication of
non-Gaussianity. There have been studies of how the three point
correlation function depends on galaxy properties \citep{kayo,jingbo,
gazta}. While the first two papers do not 
find any statistically significant effects, \citet{gazta} find
statistically significant colour and luminosity dependence. The
present work studies how the largest known non-Gaussian features, the 
filaments, depend on galaxy properties. 

This paper presents progress on many counts compared to our earlier
work \citep{pandey} based on the SDSS DR1. The earlier work was
restricted to single volume limited subsamples from each of the two
equatorial strips. The absolute magnitude range was divided into two
equal halves, and these were analyzed to test for a luminosity
dependence in the filamentarity. For a statistically significant
conclusion it is necessary to also estimate the cosmic variance
(sample to sample variation) of the filamentarity which was done by
bootstrap resampling. We found that the difference in filamentarity
between the two magnitude bins was in excess of the fluctuations
expected from cosmic variance, thereby establishing the statistical
significance.  The SDSS DR4 covers a large contiguous region in the
Northern Galactic Cap with a few gaps still to be covered.  The SDSS
has been carried out in overlapping strips of $5^{\circ}$ width
oriented along great circles. Since the detection of coherent
large-scale features requires a completely sampled region with no
gaps, we have extracted seven non-overlapping strips of width
$2^{\circ}$ each. The volume corresponding to each strip is nearly two
dimensional and we have collapsed the thickness and analysed the
resulting 2-D galaxy distribution. The increased number of strips
provide a better estimate of the cosmic variance leading to more
robust conclusions. The analysis has been carried out on five
overlapping bins in absolute magnitude. We have extracted separate
volume limited subsamples corresponding to each magnitude bin and
compared the filamentarity across them.  The earlier analysis has been
extended here to also consider the colour and morphology dependence,
which has been tested in each of the volume limited subsamples. 

Here we would like to mention that the statistics that we are using to
quantify the filamentarity is not absolute in the sense that it is
sensitive to the volume and galaxy number density of the sample.
It is thus only meaningful to compare different galaxy samples
provided they have the same volume (identical shape and size) and
number density. This is an important consideration when constructing
the galaxy samples for which we test the luminosity, colour and
morphology dependence. 

 A brief outline of our paper follows. Section 2 describes the data and
 Section 3 the method of analysis. Our results for the SDSS data are
 presented in Section 4, in Section 5 we compare our results with the
 predictions of a semi analytic model of galaxy formation and finally
 we present our conclusions in Section 6.

It may be noted that  we have used a $\Lambda$CDM cosmological model with
 $\Omega_{m0}=0.3$,  $\Omega_{\Lambda0}=0.7$ and $h=1$ throughout. 

\section{Data }

The SDSS is an imaging and spectroscopic survey of the sky
(\citealt{york}) in five photometric bandpasses, u,g,r,i,z with
effective wavelengths of $3540 \rm A^\circ$, $4760 \rm A^\circ$, $6280
\rm A^\circ$, $7690 \rm A^\circ$ and $9250 \rm A^\circ$ respectively
(\citealt{fuku}, \citealt{smith}) to a limiting r band magnitude $\sim
22.2$ with $95\%$ completeness. A series of pipelines that perform
astrometric calibration (\citealt{pier}), photometric reduction
(\citealt{lup}) and photometric calibration (\citealt{hog}) are used
to process the imaging data and then objects are selected from the
imaging data for spectroscopy.  In the SDSS
DR4 (\citealt{adel}) the imaging data covers 6670 $deg^2$ of the sky
whereas the spectroscopy covers a total area of 4783 $deg^2$. The
spectroscopic data includes 673,280 spectra with approximately 480,000
galaxies, 64,000 quasars and 89,000 stars.  The survey area covers a
single contiguous region in the Northern Galactic Cap and three
non-contiguous region in the Southern Galactic Cap. Our analysis is
restricted to the Northern Galactic Cap region only.

\begin{figure}
\rotatebox{0}{\scalebox{.3}{\includegraphics{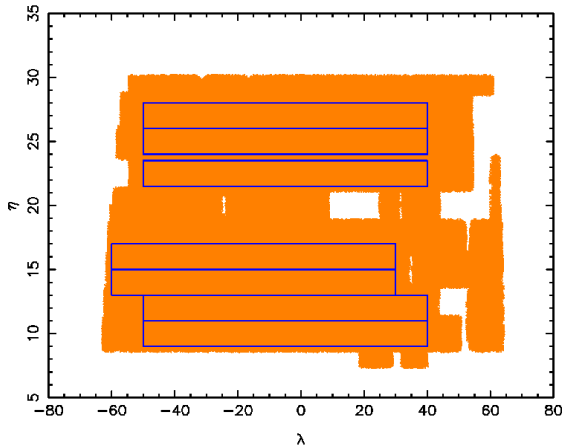}}}
\caption{The shaded part shows the contiguous region covered by the
  seven overlapping strips constituting our primary data. The bounded
  regions within show the seven $2^\circ$ wide non-overlapping strips
  used in our analysis.  }
\label{fig:1}
\end{figure}

The samples analyzed here were obtained from the SDSS DR4 Skyserver\footnote{
http://cas.sdss.org/dr4/en/}. This is a web interface to the SDSS
data archive server. The SDSS surveys the sky in overlapping stripes
\footnote{http://www.sdss.org/dr4/coverage/atStripeDef.par} of width
 $5^{\circ}$ which form great circles on the sky. The stripes are most
 conveniently described using survey co-ordinates ($\lambda, \eta$)
 defined in \citet{stout}. Lines of constant $\eta$ are great circles
 aligned with the stripes and lines of constant $\lambda$ are are
 small circles along the width of the stripes. Each stripe is centered
 along a line of constant $\eta$ separated from the adjoining stripe
 by $2.5^{\circ}$. We have downloaded data from seven overlapping
 stripes which form a nearly contiguous region shown in Figure
 \ref{fig:1}.  We have selected all objects identified as galaxies
 with extinction corrected r band Petrosian magnitude $r<17.77$ over
 the redshift range $0.01 \geq z \geq 0.2$.

The primary data is not contiguous and has a few gaps.  Since our
analysis requires contiguous regions, we have extracted seven
non-overlapping strips each of width $2^\circ$ in $\eta$ and spanning
$90^\circ$ in $\lambda$ (Figure \ref{fig:1} and Table~1).  Only
galaxies with extinction corrected Petrosian r band magnitude in the
range $ 14.5 \geq m_r \geq 17.77 $ were used.  For each galaxy the
absolute magnitude was computed using its redshift, apparent
magnitude, K correction and reddening correction
(\citealt{slegel}). We adopt a polynomial fit (\citealt{park1}) for
the mean K-correction, $K(z)=2.3537(z-0.1)^2+1.04423(z-0.1)-2.5
log(1+0.1)$.

We construct volume limited sub-samples using the absolute magnitude
 bins listed in Table~2. The five overlapping absolute magnitude bins
 are referred to as ``bin 1'', ``bin 2'',... ``bin 5'' in order of
 increasing luminosity. Each bin extends over a different redshift
 range. The thickness of the volume corresponding
 to each bin increases with redshift. We have extracted regions of
 uniform thickness $6 \, h^{-1} \, {\rm Mpc}$. This 
corresponds to  $2^{\circ}$ at redshift $0.06$, and for some bins it was
 necessary to  discard the  range $z<0.06$ where the thickness was
 less than $6 \, h^{-1} \, {\rm Mpc}$.  The redshift range which we 
use for final galaxy samples that were analyzed are 
 shown in Table~2. The faintest  luminosity bin has the smallest area
 (Table~2), 
 and the area of the other bins increases monotonically with
 luminosity. A smaller magnitude range was chosen for the faintest
 bin so as to ensure that it covers a reasonably large area while
maintaining a  number density comparable to the other bins. 
The galaxy number density is maximum for bin 2 (Table~2)
which is centered at 
 $M^*$ which has a  value  $M*=-20.44\pm0.01$ \citep{blan3} for the
 SDSS. The number density falls for the other bins.
The number of galaxies for all the strips in each bin are also
 tabulated in Table~1.

\begin{table*}{}
\caption{This shows the $(\lambda,\eta)$ range  of the seven
 non-overlapping strips  extracted from the primary data. For each
 strip the table shows the number of galaxies in volume limited
 subsamples bin 1, bin 2, .. etc. with absolute magnitude and redshift
 limits  given in Table~2.}
\begin{tabular}{|c|c|c|c|c|c|c|c|}
\hline
Strip number & $\lambda(^\circ)$ & $\eta(^\circ) $ & bin 1 &
bin 2 & bin 3 & bin 4 & bin 5\\
\hline
 1 &$-50\leq\lambda\leq 40$&$9\leq\eta\leq 11$&$875$&$2016$&$2108$&$1943$&$1623$ \\
 2 &$-50\leq\lambda\leq 40$&$11\leq\eta\leq 13$&$839$&$1740$&$1780$&$1695$&$1591$ \\ 
 3 &$-60\leq\lambda\leq 30$ &$13\leq\eta\leq 15$&$663$&$1492$&$1665$&$1597$&$1540$ \\ 
 4 &$-60\leq\lambda\leq 30$&$15\leq\eta\leq 17$&$678$&$1547$&$1629$&$1569$&$1593$\\
 5 &$-50\leq\lambda\leq 40$&$21.5\leq \eta \leq 23.5$&$846$&$1878$&$1863$&$1703$&$1595$ \\
 6 &$-50\leq\lambda\leq 40$&$24\leq\eta\leq 26$&$782$&$1905$&$1854$&$1839$&$1668$\\ 
 7 & $-50\leq\lambda\leq 40$&$26\leq\eta\leq 28$&$648$&$1550$&$1573$&$1692$&$1633$ \\ 
\hline
\end{tabular}
\end{table*}

\begin{table*}{}
\caption{This shows the absolute magnitude and redshift limits for the
  different volume limited subsamples analyzed. The first three bins
  actually extend to lower redshifts, but only the region beyond
  $z=0.06$ has been kept so that it is possible to extract a volume of
  uniform thickness $6 \, h^{-1}\, {\rm Mpc}$.  The area and the
  average galaxy number density with $1-\sigma$ variations from the
  $7$ strips are also shown. The last two column shows the value
  of the galaxy colour $(u-r)_c$ and the value of the concentration
  index $c_{i,c}$ used to divide the data into equal numbers of
  red/blue and elliptical/spiral galaxies respectively.}
\begin{tabular}{|c|c|c|c|c|c|c|}
\hline bin & Absolute Magnitude range & Redshift range & Area\, [
$10^{4}\, h^{-2}\, {\rm Mpc^2}$] & Density \, [$10^{-2}\, h^{2} {\rm
Mpc}^{-2}$] & $(u-r)_c$ & $c_{i,c}$ \\ 
\hline 
bin 1 &$-19.5 \geq M_r \geq -20$ & $0.06\leq z
\leq 0.093$ & $3.37$ & $2.25 \pm 0.26$ & $2.14$ & $2.53$ \\ 
bin 2 &$-19.75 \geq M_r \geq
-21.25$ & $0.06\leq z \leq 0.103$ &$4.66$ & $3.71\pm 0.41$& $2.35$ & $2.67$ \\ 
bin 3
&$-20\geq M_r \geq -21.5$ & $0.06 \leq z \leq 0.114$ & $6.22$ & $2.86
\pm0.27$ & $2.41$ & $2.72$ \\ 
bin 4 &$-20.25 \geq M_r \geq -21.75$ & $0.06\leq z \leq
0.126$ & $8.08$ & $2.13 \pm 0.15$ & $2.45$ & $2.75$ \\ 
bin 5 &$-20.5 \geq M_r \geq -22$ &
$0.067\leq z \leq 0.14$  & $9.87$ &$1.63 \pm 0.037 $ & $2.5$  & $2.78$\\ \hline
\end{tabular}
\end{table*}

\begin{figure*}
\rotatebox{-90}{\scalebox{.8}{\includegraphics{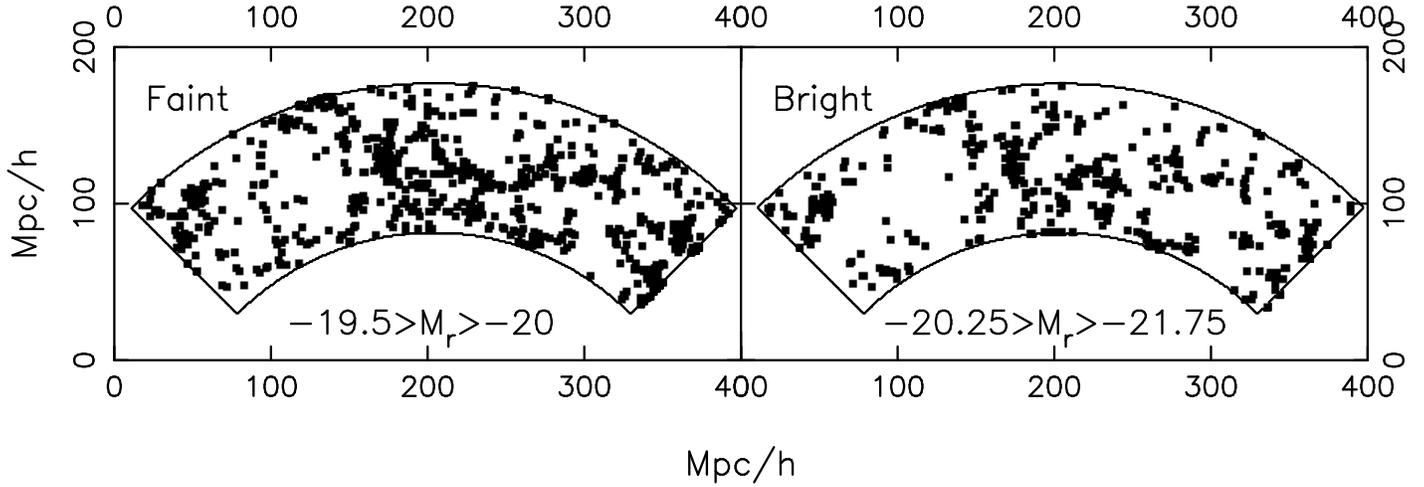}}}
\caption{ This shows the distribution of faint and bright galaxies in
  one of the strips in magnitude bin 1 and bin 4 respectively after
  three round of coarse-graining.  }
\label{fig:2}
\end{figure*}

Studies using N-body simulations show that the statistics that we use
to characterize the filamentarity in the galaxy distribution is
sensitive to the number density as well as the shape and size of the
volume of the sample.  To test for luminosity dependence it is
necessary to compare different luminosity bins which as seen in
Table~2 have different number densities and cover different areas. To
ensure that these factors do not influence our results, we extract
regions of identical shape and size as bin 1 from the other bins and
use these to test for luminosity dependence.  Tests using dark matter
$\Lambda$CDM N-body simulations show that galaxy number density
variations of $\sim 50 \%$ do not produce statistically significant
effect on the filamentarity (Figure \ref{fig:11}). We have discarded
randomly chosen galaxies from bins 2 and 3 so as to make the galaxy
number density of each strip of these bins exactly equal to the mean
density of bin 1. The galaxy number density of bin 4 is within $10 \%$
of that of bin 1 while it is $27\%$ lower in bin 5. Since these are
lower than bin 1, and as they are within the permissible range of
variation we do not dilute the galaxy number density in bins 4 and 5.
Figure \ref{fig:2} shows the galaxy distribution in two different
luminosity bins drawn from the same strip. It is to be noted that the
samples with reduced area and diluted galaxy number density are used
only for testing luminosity dependence.

\begin{figure*}
\rotatebox{-90}{\scalebox{.8}{\includegraphics{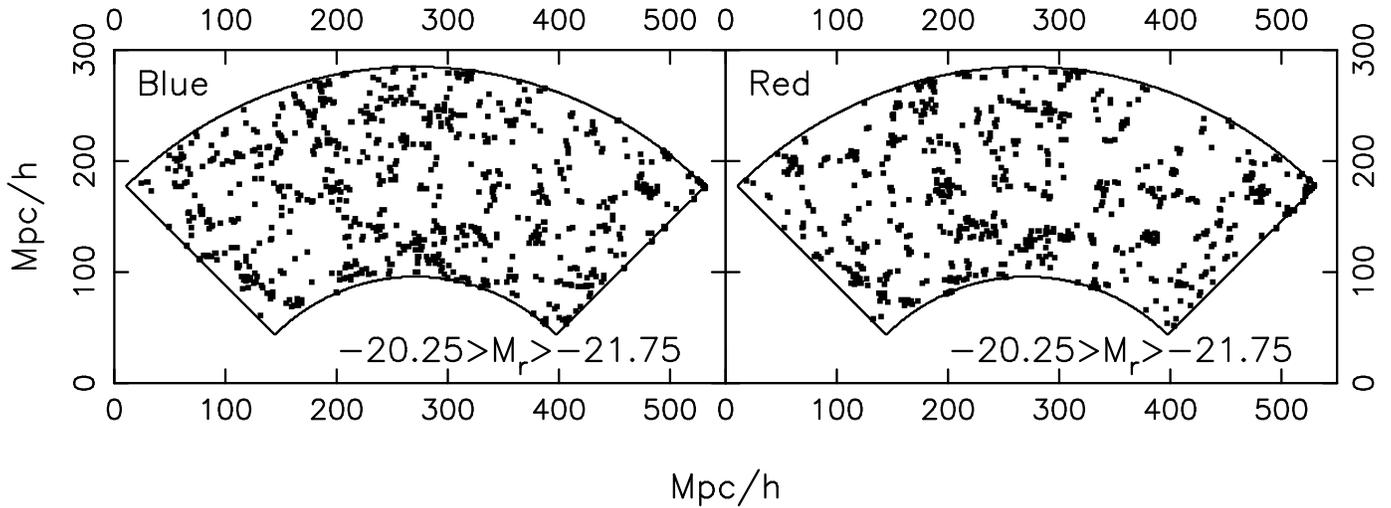}}}
\caption{ This shows the distribution of blue and red galaxies in one
  of the strips in magnitude bin 4 after three round of coarse-graining.  }
\label{fig:3}
\end{figure*}

\begin{figure*}
\rotatebox{-90}{\scalebox{.8}{\includegraphics{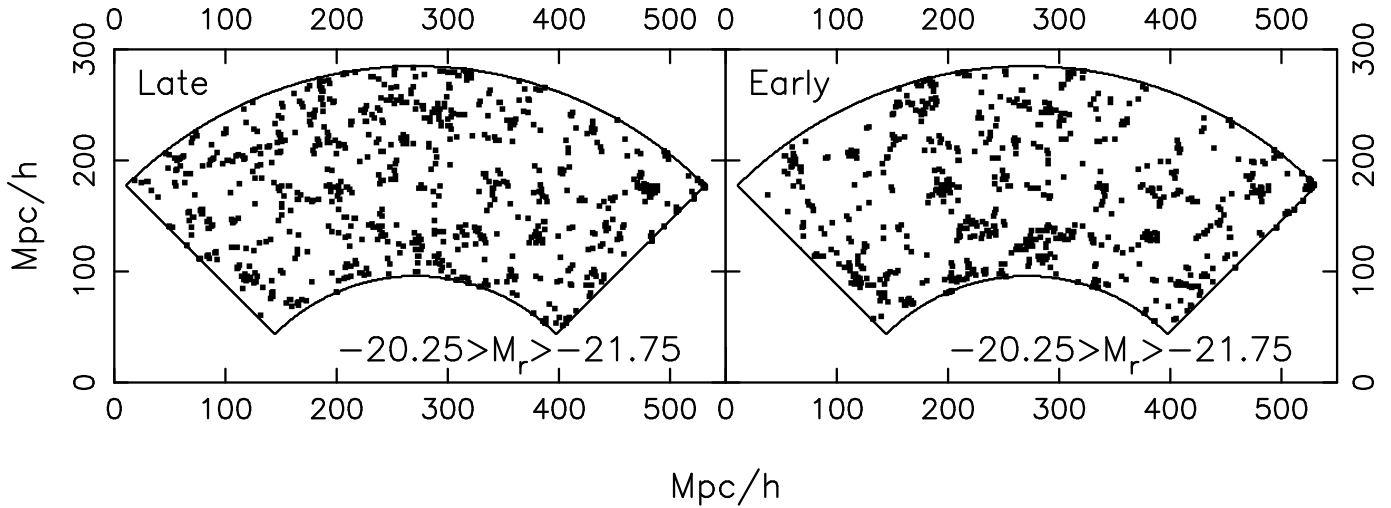}}}
\caption{This shows the distribution of early and late type galaxies
  in one of the strips in magnitude bin 4 after three round of
  coarse-graining.  }
\label{fig:4}
\end{figure*}

We separately study the colour and morphology dependence in each of
the magnitude bins listed in Table~1 and Table~2. We do not compare
the colour and morphology dependence across different  luminosity
bins. When testing for colour dependence in a fixed luminosity bin,
all the galaxies  (Table~1) in the entire area of the bin (Table~2)
are classified as either red or blue galaxies. 
The galaxy $u-r$ colours are known to have a bimodal distribution
\citep{strat}. In our analysis  we determine  a value  
$(u-r)_c$ for the colour such that it divides each magnitude bin into
equal number of red (i.e. $u-r>(u-r)_c$) and blue (i.e. $u-r\leq(u-r)_c$)
galaxies. As a consequence the number density of red and blue galaxies
in each luminosity bin are exactly equal and have a value half that of
the density shown in Table~2. 
 The value of $(u-r)_c$ varies slightly across the luminosity bins, and
the values are listed in Table~2. It may be noted that \citet{strat}
finds that a color selection criteria $(u-r)_c=2.22$ ensures that the
red galaxies are 'ellipticals' with $90 \%$ completeness. Figure
\ref{fig:3} shows the distribution of red and blue galaxies in one of
the subsamples.

 The morphological classification was carried out using the
 concentration index defined as $c_i=r_{90}/r_{50}$ where $r_{90}$ and
 $r_{50}$ are the radii containing $90 \%$ and $50 \%$ of the
 Petrosian flux respectively. This has been found to be one of the
 best parameter to classify galaxy morphology (\citealt{sima}).
 Ellipticals are expected to have a larger concentration indices than
 spirals. It was found that $c_i \simeq 3.33$ for a pure
 de-Vaucouleurs profile (\citealt{blan2}) while $c_i\simeq 2.3$ for a
 pure exponential profile (\citealt{strat}). \citet{strat} show that
 using a cutoff $c_{i,c}=2.6$ divides the sample into ellipticals and
 spirals at $83 \%$ completeness. For each luminosity bin we have
 chosen a cutoff $c_{i,c}$ that partitions the galaxies into two equal
 halves, one predominantly ellipticals and the other spirals.  The
 values of $c_{i,c}$ varies across luminosity bins and are listed in
 Table~2. Figure \ref{fig:4} shows the distribution of early and late
 type galaxies in one of the subsamples.

Finally we note that there is an incompleteness in the SDSS survey
arising from a restriction which prevents the redshift of two galaxies
at a very small angular separation to be measured. This incompleteness
is not expected to introduce a luminosity, colour or morphology
dependence and hence we do not take this into account. 

\section{Method of analysis}
Each of the subsamples described in the previous subsection have a
much greater (16-35 times) linear dimension compared to their
thickness and so can be reasonably treated as two dimensional. The 2D
galaxy distribution was embedded in a $1 \, h^{-1} {\rm Mpc} \times 1
\,h^{-1} {\rm Mpc}$ 2D rectangular grid. The galaxy distribution was
represented as a set of 1s and 0s on a 2-D rectangular grid by
assigning the values 1 and 0 to the filled and empty cells
respectively. Note that $\sim 10 \%$ of the filled cells have 2
galaxies while  the number is substantially smaller for 3 galaxies or
more, and hence  the 1s and 0s provide a fair representation of the
galaxy distribution on the length-scales of interest. 
The grid cells which are beyond the boundaries of the
survey were assigned a negative value in order to distinguish them
from the empty cells within the survey area.

\begin{figure}
\rotatebox{-90}{\scalebox{.35}{\includegraphics{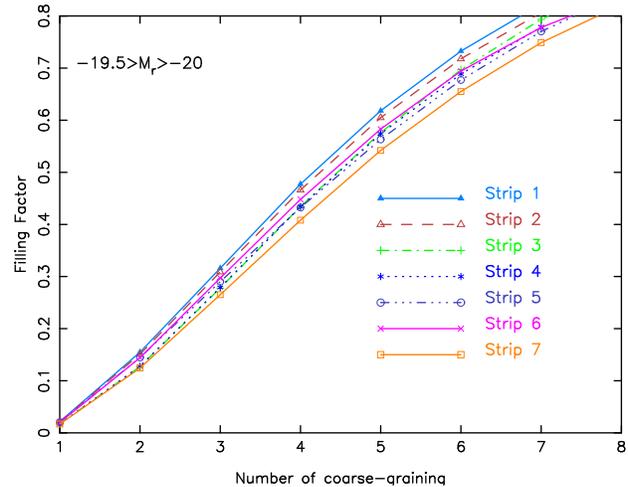}}}
\caption{This shows the Filling Factor ($FF$) as a function of
 the number of iterations of coarse-graining applied to the initial galaxy
 distribution for seven SDSS different strips in the luminosity bin
 indicated in the figure.}
\label{fig:5}
\end{figure}

The next step is to use an objective criteria to identify the coherent
large-scale structures visible in the galaxy distribution.  We use a
``friends-of-friends''(FOF) algorithm to identify interconnected
regions of filled cells which we refer to as clusters. In this
algorithm any two adjacent filled cells are referred to as friends.
Clusters are defined through the stipulation that any friend of my
friend is my friend. The distribution of 1s on the grid is very sparse
with only $\sim 1 \%$ of the cells being filled.  Also, the filled
cells are mostly isolated, and the clusters identified using FOF,
which contain only a few cells each, do not resemble the large-scale
coherent structures seen in the SDSS strips. It is necessary to
coarse-grain the galaxy distribution so that the large scale
structures may be objectively identified. 
In every iteration of coarse-graining we fill up all the 
empty cells adjacent to every filled cells (i.e. cells at the 4 sides
and 4 corners of a filled cell), causing every filled cell to grow
fatter. The size of an isolated filled cell is $(2 N+1)
\times (2 N+1)$ after  $N$ iterations of coarse-graining and for
example it is $7 \, h^{-1}\, {\rm Mpc} \times 7 \, h^{-1}\, {\rm Mpc}$
after $3$ iterations. 
 This causes clusters to 
grow, first because of the growth of filled cells, and then by the
merger of adjacent clusters as they overlap.  Increasing the 
coarse-graining further induces  the percolation transition when a
large 
fraction of the clusters connect up into a single, multiply connected
network of filaments encircling voids.  As coarse-graining proceeds
even further the network grows to some extent, and eventually the
elements become very thick filling up the entire survey region
washing away any visible  pattern. The filling factor $\FF$,
defined as the fraction of 
cells within the survey area that are filled, {\it ie.}
\begin{equation}
\FF=\frac{{\rm Total \, No.\,  of\,  Filled \, Cells}}{{\rm Total \,
 No.\,  of \, Cells \, Inside \, the \,  
 Survey \, Area}}
\end{equation}
 increases from $\FF \sim$ 0.01 to $\FF =$ 1 as the coarse-graining is
 increased.  The filling factor is around $FF \sim 0.3$ after $3$
 rounds of coarse-graining (Figure \ref{fig:5}). 
 So as to not restrict our analysis to an arbitrarily chosen
 level of coarse-graining, we analyze the clusters identified in the
 pattern  of 1s and 0s after each iteration of coarse-graining.  
Earlier studies using the SDSS DR1 \citep{pandey} show   that  
the  percolation transition occurs in the $FF$ range $0.5-0.6$.

The geometry and topology of a two dimensional cluster can be
described by the three Minkowski functionals, namely its area $S$,
perimeter $P$, and number of holes or genus $G$. We have tested that
the cluster area $S$ is proportional to the actual number of 
galaxies  within the cluster boundary. This holds at each level of
coarse-graining, with the proportionality constant increasing with
coarse-graining.  
The ratio $T=S/P$
characterizes the thickness of the cluster, and $P$  the  extent of its
boundary. Since the clusters are largely filamentary (as we shall see
later),  we may interpret $P$ as an estimate of the length of the
cluster. 
The ratio $P/G$ has dimensions of length and it is
particularly significant after the onset of percolation \citep{bharad1}
 as it
characterizes the  length of the cluster boundary per void. 
It is possible to quantify the shape of
the cluster using a single 2D ``Shapefinder'' statistic (Bharadwaj et
al. 2000) which is defined as the dimensionless ratio
\begin{equation}
{\cal F}=\frac{P^2 - 4 \pi S}{P^2 + 4 \pi S}\,, 
\end{equation}

 which by construction has values in the range $0 \le {\cal F} \le
  1$. It can be verified that ${\cal F} =1$ for an ideal filament
  which has a finite length and zero width, whereby it subtends no
  area ($S=0$) but has a finite perimeter ($P>0$). It can be further
  checked that ${\cal F}=0$ for a circular disk, and intermediate
  values of ${\cal F}$ quantifies the degree of filamentarity with the
  value increasing as a cluster is deformed from a circular disk to a
  thin filament.

The definition of ${\cal F}$ needs to be modified when working on a
rectangular  grid of spacing $l$.  An ideal filament, represented on a
grid,   has the minimum possible width {\it i.e.} $l$,  and its perimeter
$P$ and area $S$ are related as 
$2 S=(P- 2 l)l$. At the other extreme we have  $P^2=16 S$ for a square 
shaped cluster on the grid. We introduce the  2D Shapefinder statistic 

\begin{equation}
{\cal F} = \frac{(P^2 - 16 S)}{(P-4 l)^2}
\end{equation}

to quantify the shape of  clusters on a grid.  By definition 0$\le
{\cal F} \le$ 1.  ${\cal F}$ quantifies the degree of filamentarity of the 
cluster, with ${\cal F}$ = 1 indicating a filament  and ${\cal F}$ =
0, a square, and ${\cal F}$ changes from $0$ to $1$ as a square is
deformed to a filament.  

At every stage of coarse-graining we have a number of clusters, each
of which is characterized by its shape ${\cal F}_i$  and size  $S_i$. 
It is desirable to identify a single statistical
quantity to capture the filamentarity of the entire set of clusters.  
We have considered two possibilities $F_1$ and $F_2$, respectively the
first and second area weighted moments of the filamentarity.  
Our aim being to quantitatively compare the filamentarity of different
galaxy samples and draw statistically significant conclusions, we
find that it is more advantageous to use $F_2$ as a statistical
discriminator. Though both $F_1$ and $F_2$ show  similar behaviour,
the differences in $F_2$ between point distributions  with different 
clustering properties is more pronounced as compared to $F_1$
\citep{bharad1}.  The average filamentarity $F_2$
is defined as the mean filamentarity of all the clusters in a slice
weighted by the square of the area of each clusters

\begin{equation} 
F_2 = {\sum_{i} {\cal S}_i^2 {\cal F}_i\over\sum_{i}{\cal S}_i^2} \,. 
\label{eq:a6}
\end{equation}

In the current analysis, we study  the average filamentarity $F_2$ as
a function of $FF$ to quantify the degree of filamentarity in each of  
the SDSS subsamples and investigate how the average filamentarity $F_2$
changes with intrinsic galaxy properties like luminosity, color and
morphology. We also use the area weighted second moment (defined
analogous to eq. (\ref{eq:a6}))  of $S/P$, $P$ and $P/G$  to
quantify the average cluster thickness, length and length per 
void respectively. 

At a specified level of coarse-graining the value of the filling
factor ($FF$) has small differences from sample to sample as seen in
Figure \ref{fig:5}.  This makes it difficult to compare the
filamentarity in different samples using the values of the average
filamentarity ($F_2$) as a function of $FF$ (Figure \ref{fig:6}). We
overcome this by interpolating the values of $F_2$ as a function of
$FF$ at uniformly chosen values which are same for all the
samples. The average thickness and length are also interpolated and
analyzed in a similar fashion.
 Further, since the later stages of coarse graining tend to
wash out the features of the galaxy distribution, we have restricted
our analysis to the range of filling factor $0 \le FF \le 0.75$.

\section{Results}

\begin{figure}
\rotatebox{-90}{\scalebox{.35}{\includegraphics{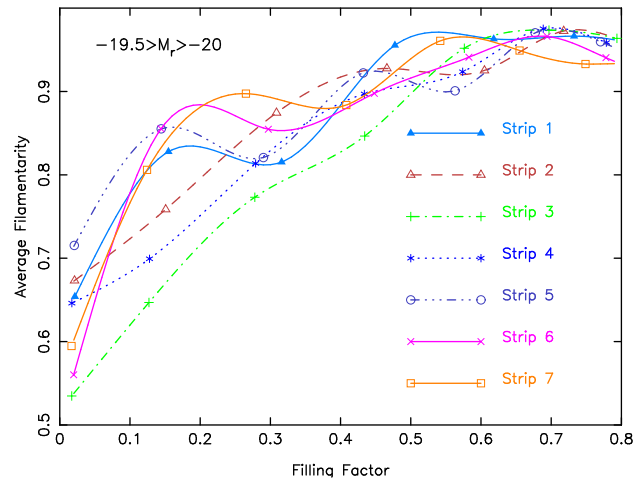}}}
\caption{This shows the average filamentarity ($F_2$) as a function of
 filling factors ($FF$) for all seven individual SDSS strips in the luminosity
 bin indicated in the figure. The markers show the values for each
 iteration of coarse-graining, whereas the curves show the
 interpolation used to determine $F_2$ at equally spaced values of $FF$.}
\label{fig:6}
\end{figure}

\begin{figure*}
\rotatebox{-90}{\scalebox{.75}{\includegraphics{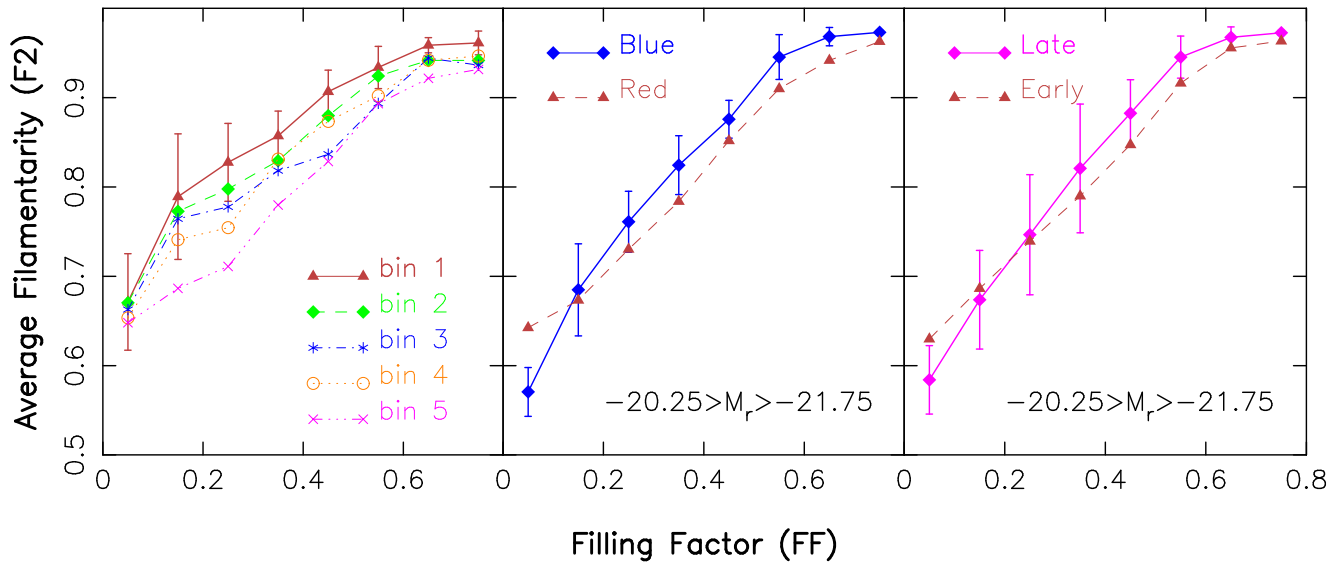}}}
\caption{The average filamentarity ($F_2$) as a
  function of filling factor ($FF$) for the galaxies divided by
  luminosity (left), colour (middle) and morphology (right). The
  $1-\sigma$ errorbars are shown only for a single curve in each
  panel, the other curve(s) shown in the same panel have similar
  errorbars. The different magnitude bins used in studying  the
  luminosity dependence (Table 2.) are shown in the left panel. The
  colour and 
  morphology dependence  are shown for a single luminosity bin
  (bin 4 of Table 2.), the other luminosity bins show similar colour
  and morphology dependence.}
\label{fig:7}
\end{figure*}

We first study how the average filamentarity varies across the
different luminosity bins discussed in Section 2. For each luminosity
bin we have seven realisations of the galaxy distributions
from the seven non-overlapping strips. The results from all the 
individual strips in luminosity bin 1 are shown together in Figure 
\ref{fig:6}. For each bin we use the results from these seven samples
to determine the mean and variance of $F_2$ at uniformly chosen values
of $FF$. The results are shown in the left panel of 
Figure \ref{fig:7}. We find that for
a fixed value of $FF$, the value of $F_2$ decreases with increasing
luminosity. We note that there is some deviation from this behaviour
at $FF>0.4$ between bins 3 and 4, but the overall trend is still valid
when we compare these bins with fainter or brighter samples. Further,
the luminosity dependence is enhanced with increasing luminosity.

At each values of $FF$ we use the Student's t-test to quantify the
statistical 
significance of the difference in the mean $F_2$ between different luminosity
bins. The variance of $F_2$  is very similar  in the different
luminosity bins and we estimate the standard error for the difference
in the means using 

\begin{figure}
\rotatebox{-90}{\scalebox{.35}{\includegraphics{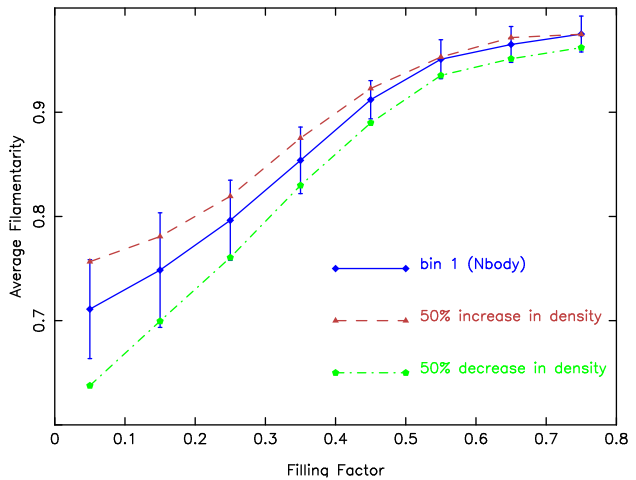}}}
\caption{This shows how the average filamentarity ($F_2$) as a
 function of filling factors ($FF$) changes if the galaxy number
 density is changed by $50 \%$. Each curve shows results from 18
 strips identical in area and thickness as bin 1 (Table~2) drawn from
 3 independent $\Lambda$CDM dark matter N-body simulation. The solid
 curves and the errorbars are for mock strips whose galaxy number
 density exactly matches the mean density of bin 1 (Table~2).}
\label{fig:11}
\end{figure}

$s_D=\sqrt{\frac{\sum_{i\in A} (x_i-\bar x_A)^2 +\sum_{i\in B}
     (x_i-\bar x_B)^2}{N_A+N_B-2}(\frac{1}{N_A}+\frac{1}{N_B})}$ 

where the sum is over the points in the two samples A and B which are
being compared. Here $\bar x_A$ and $\bar x_B$ referes to the mean,
and $N_A$ and $N_B$ refers to  the number of data  points. In our case
$N_A=N_B=7$ . We use $t=\frac{\bar   x_A-\bar x_B}{s_D}$ to estimate 
the significance of the differences in the means. This is expected to
follow a Student's t-distribution with $12$ degrees of freedom. 
We accept the difference in the means as being statistically
significant if the probability of its occurring by chance is less than
$5 \%$. We find that there are no statistically significant differences
between adjacent luminosity bins except  the  two brightest ones (bins
4 and 5). All the non-adjacent bins (eg. bin 1 and bin 3, etc.) show
statistically significant differences at most values of $FF$.  As
noted earlier, the $F_2$ curves are sensitive to the galaxy number
density. While bins 2 and 3 have been culled so that they have the
same number density as bin 1, bins 4 and 5 have  number densities 
which are $10 \%$ and $27 \%$ lower respectively. Figure \ref{fig:11}
shows the effect of varying the number density by $50 \%$ in  18
independent realizations of  mock galaxy  samples identical to bin 1 in
thickness, area and number density (Table~2) drawn from dark matter
$\Lambda$CDM  N-body simulations \citep{bharad3}. The test shows that
$50 \%$ variations in the number density do not introduce a
statistically significant effect on the filamentarity except at
$FF=0.05$. This is quite distinct from the luminosity
dependence seen in the left panel of Figure \ref{fig:7} where there
are statistically significant differences among the different
luminosity bins at all $FF$ except for $FF=0.05$.  This clearly
establishes the luminosity dependence to be genuine and not an
artifact due to number density variations. 

 We have separately analyzed the filamentarity of the red and blue
  galaxies in each luminosity bin. The results for bin 4 are shown in
  the middle panel of Figure \ref{fig:7}. We find that there is a
  statistically significant colour dependence.  At the smallest value
  of $FF$ the red galaxies exhibit a higher filamentarity than the
  blue ones.  There is a cross-over at $FF \sim 0.2$, and the blue
  galaxies have a higher filamentarity than the red ones at larger
  values of $FF$. A similar trend is found in all the other luminosity
  bins, results for which are not shown here.

The morphology dependence has been studied separately for each
luminosity bin. The results for bin 4 are shown in the right panel of
Figure \ref{fig:7}.
The ellipticals have a higher $F_2$ as compared to spirals for $FF
\leq 0.25$ whereas the spirals have a higher $F_2$ at larger  values
of $FF$.  We find that the differences are statistically significant
for $FF \leq 0.1$ and $FF \geq 0.4$. The results are similar for the
other luminosity bins not shown here. 

\begin{figure*}
\rotatebox{-90}{\scalebox{.75}{\includegraphics{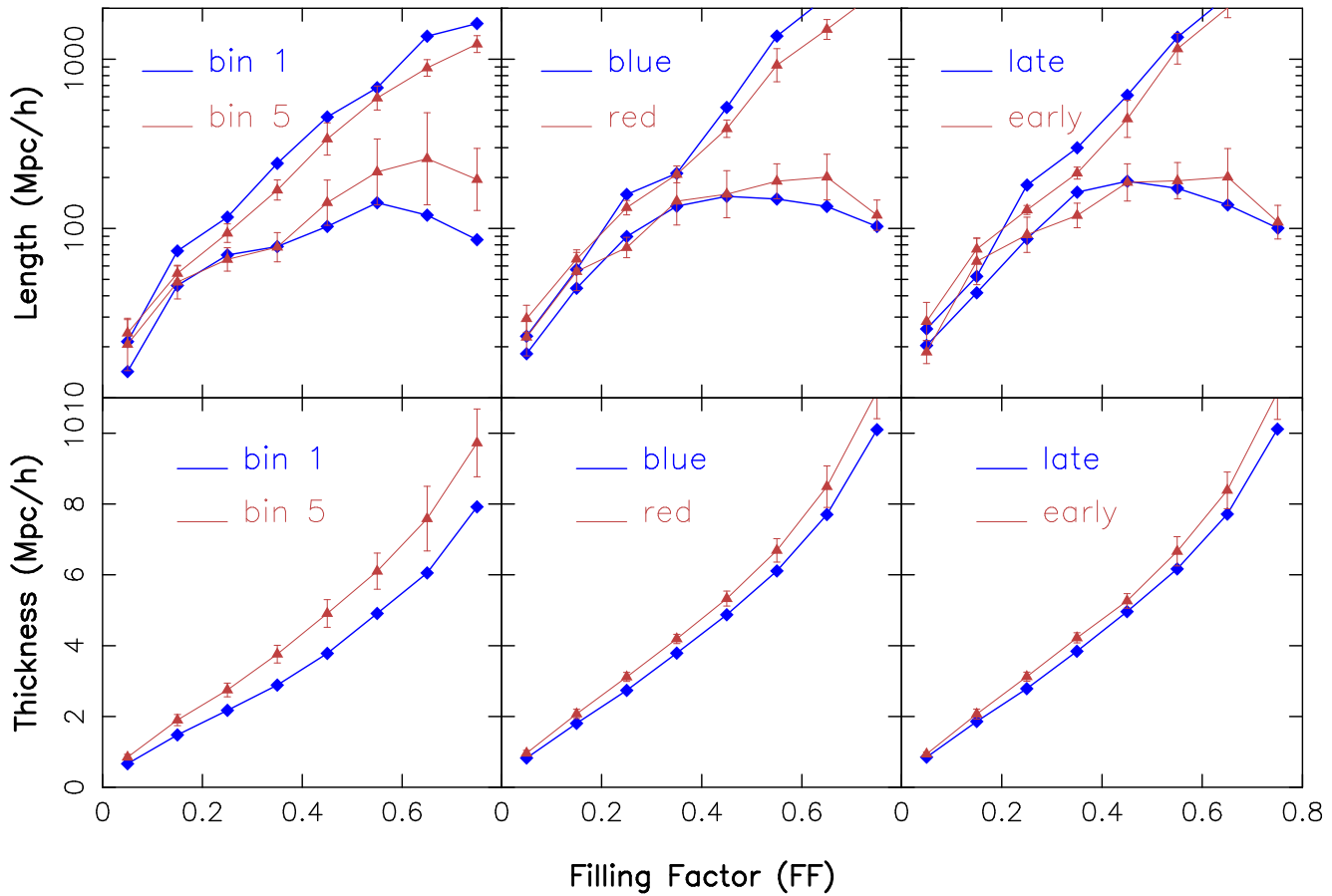}}}
\caption{The  upper panels show  the average cluster length
 and length per void while the  lower panels show the average
 cluster thickness.  The length increases nearly
 monotonically with $FF$ whereas  the length per void is stable after the
 onset of percolation at $FF\sim 0.5$. The length per void values have
 been  divided by factor of $2$ for convenience of plotting as the
 curves overlap otherwise.
 The left panels show the
 brightest and faintest luminosity bins, while the central and right
 panels show the colour and morphology dependence respectively for a
 fixed  luminosity bin (bin 4). Note that  in each panel we have shown
 $1-\sigma$ errorbars for one of the curves only, the  other curve
 has similar errorbars.}
\label{fig:9}
\end{figure*}

Figure \ref{fig:9} shows the average cluster length, length per void and 
thickness as a function of  $FF$ for two different luminosity bins,
and also  
the colour and morphology dependence for a fixed luminosity bin (bin
4). We see that the clusters are longer for the faint galaxies, blues
galaxies and the spirals as compared to the bright galaxies, red
galaxies and ellipticals respectively.  
Unlike the length which increases nearly monotonically with
coarse-graining, the cluster length  per void  is found to be  quite
stable to  coarse-graining after the onset of percolation.
 The luminosity, colour and morphology dependence found in the  cluster
 length is exactly reversed when the length per void is considered. We
 also note that the luminosity, colour and morphology dependence of
 the length and the length per void are not at a high level of
 statistical  significance, except at a few points where the data 
  are seen to lie outside the  $1-\sigma$ errorbars.  The average
  cluster thickness, on the other hand, 
  shows statistically significant differences at all values of the
  filling factor. At the same value of $FF$, the clusters are thicker
  for  the bright galaxies, red galaxies and elliptical galaxies as compared
  to their faint,   blue and spiral counterparts  respectively.  

\section{Comparison with a semi analytic model of galaxy formation}

Semi analytic models have, over the last two decades emerged as a very
powerful tool for studying galaxy formation (\citealt{white2},
\citealt{lacey1}, \citealt{kofman1}, \citealt{kofman4},
\citealt{lacey}, \citealt{cole}, \citealt{kofman2}, \citealt{kofman3},
\citealt{bagh}, \citealt{somervil}, \citealt{somervil1},
\citealt{kofman5}, \citealt{cole1}, \citealt{benson1},
\citealt{springel}). These models describe the evolution of galaxies
in a hierarchical clustering scenario incorporating all relevant physics
of galaxy formation processes (eg. gas cooling, star formation,
supernovea feedback, metal enrichment, merging etc) often in an
approximate and adhoc fashion. The detailed physics of star formation
and it's regulation by different feedback mechanisms is still poorly
understood. These models serve as simplified simulations of the galaxy
formation processes. The output of the semi analytic models are
statistical predictions of galaxy properties at some epoch and the
precision of these predictions are directly related to the accuracy of
the input physics. In this Section we investigate the luminosity, and
colour dependence of galaxy filaments in a particular semi-analytic
model and  compare the findings with those from the SDSS. We have used
the semi analytic galaxy catalogs from 
the Millennium Run simulation (\citealt{springel}), one of the largest
simulation of the growth and evolution of cosmic structures in the
universe. The details of the simulation can be found in
\citet{springel}. The physical treatment of the galaxy formation
processes in this model are described in \citet{croton}. The spectra
and magnitude of the model galaxies were computed using population
synthesis models of \citet{brujual} and we use the catalog where the
galaxy magnitudes are available in SDSS u,g,r,i,z filters.  The
catalog contains  about 9 million galaxies in the full simulation
box.  We use these to construct   mock strips from the simulation
which are otherwise identical to the SDSS strips which were analyzed. 

\begin{figure*}
\rotatebox{-90}{\scalebox{.75}{\includegraphics{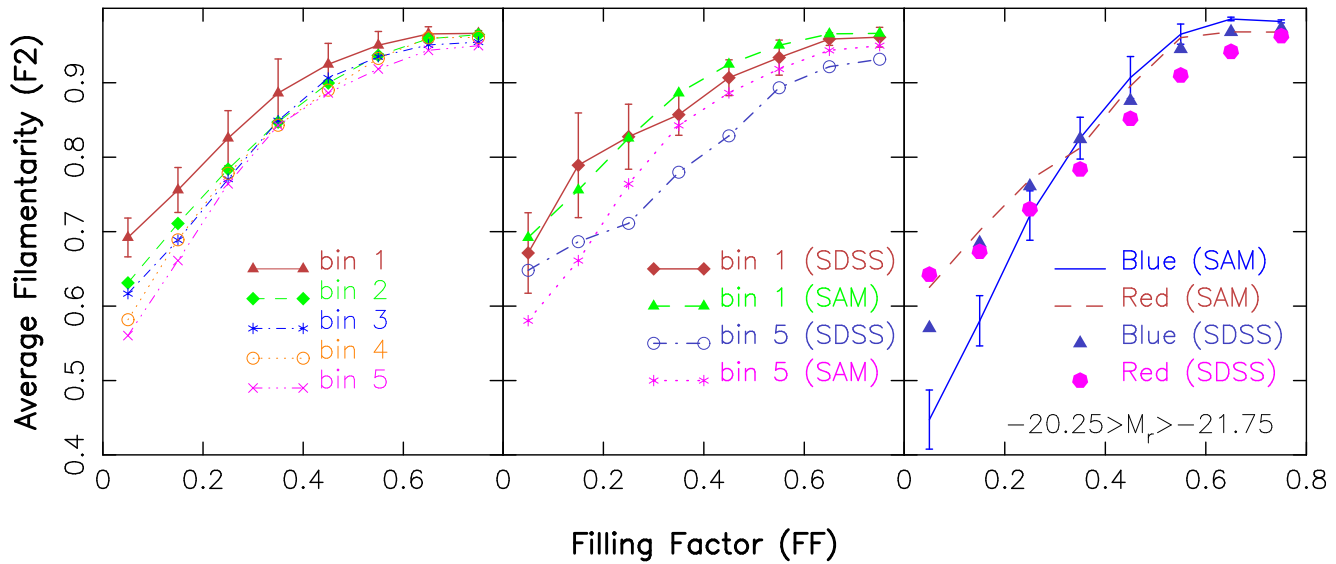}}}
\caption{ The average filamentarity ($F_2$) as a function of the
  filling  factor ($FF$) for the galaxies in the Semi-Analytic
  Model (Millennium Run). The left panel shows the luminosity
  dependence. The middle panel shows the faintest and brightest
  luminosity bins for both the actual data (SDSS) and the
  Semi-Analytic Model (SAM).  The right panel shows the colour
  dependence for a single luminosity bin (bin 4 of Table 2.).   
  The $1-\sigma$ errorbars are shown only for a single curve in each
  panel, the other curve(s)  in the same panel have similar
  errorbars. }
\label{fig:8}
\end{figure*}

The luminosity dependence in the semi-analytic model is shown in the
left panel of Figure \ref{fig:8}. This is found to exhibit a strong
luminosity dependence. Although the behaviour is qualitatively similar
to the actual data, the luminosity dependence is noticeably less
pronounced for the brighter galaxies in the semi-analytic model. The
differences between the behaviour in the actual data and the
semi-analytic model are clearly seen in the middle panel of Figure
\ref{fig:8} which, for two luminosity bins, shows the results for the
actual data and the semi-analytic model together. Note that while
there is a reasonable agreement for the faint luminosity bins, there
are sharp differences between the data and the simulations in the
filamentarity of the high luminosity galaxies.

The right panel of Figure \ref{fig:8} shows the colour dependence of
the filamentarity in the semi-analytic model. Here again, although the
model is qualitatively similar to the data, the colour dependence
predicted in the model is substantially in excess of those seen in the
actual data. The differences between the data and the simulation is
particularly noticeable at low $FF$ where the filamentarity of the
blue galaxies in the model is found to be substantially below that of
the actual data. The simulation results for the red galaxies are in
rough agreement with the SDSS data.

The concentration index ($c_i$) was not directly available in  the
particular simulated catalogue which we have used and hence we did not
study the morphology dependence for  the semi-analytic model. 

\section{Summary  and Conclusion}
Filaments are the largest known statistically significant coherent
features visible in the galaxy distribution. We test if the degree of
filamentarity depends on the galaxy properties.   We find evidence for 
statistically significant luminosity, colour and morphology dependence
in the average filamentarity $F_2$ studied as a function of the
filling factor $FF$.

Comparing the average filamentarity $F_2$ of different luminosity bins
(Figure \ref{fig:7}), we find that the fainter galaxies have a more
filamentary distribution as compared to the brighter ones.  These
differences exist at nearly all values of $FF$.  The drop in
filamentarity with increasing luminosity is found to be particularly
more pronounced in the two brightest luminosity bins which we have
considered, both of which are brighter than $M^*$.  Comparing the
thickness and length (Figure \ref{fig:9}), we find that at a fixed
value of $FF$ structures traced by the bright galaxies are thicker and
shorter than those traced by the faint ones. While the differences in
the length are statistically significant only at large $FF$, the
differences in the thickness are statistically significant at all
values of the filling factor. With increasing coarse-graining there is
a percolation transition at $FF \sim 0.5$ beyond which there is an
interconnected network of filaments encircling voids. After the onset
of percolation the length per void gives an estimate of the
circumference of the filaments encircling the voids and dividing this
by $\pi$ gives an estimate of the average void diameter. We find that
the faint galaxy distribution is more porous (larger number of holes
or voids) and has smaller voids with a typical void diameter $\sim 80
\pm 62 \, h^{-1} \, {\rm Mpc}$ as compared to the bright galaxy
distribution which has a typical void diameter $\sim 165 \pm 140 \,
h^{-1} \, {\rm Mpc}$.  Note that these estimates of the void diameter
could be somewhat flawed because the estimated void diameters exceed
the radial extent ($96 \, h^{-1}\, {\rm Mpc}$) of the samples which we
have used to test luminosity dependence (Figure \ref{fig:2}), but they
serve the purpose of demonstrating the luminosity dependence of the
void size. Our findings indicate that galaxies of
different luminosity are not uniformly distributed along the cosmic
web. The brighter galaxies are preferentially distributed in more
compact and thicker regions with large voids in between whereas the
fainter galaxies inhabit thin, elongated regions with more numerous
voids of smaller diameter.  We note that studies using the projected
two-point correlation function (eg. \citealt{nor1}; \citealt{zehavi})
show an increase in correlation amplitude with increasing luminosity,
the effect being pronounced beyond $L^*$. These earlier results are
consistent with our findings but they do not have any information
about the shapes of the clustering patterns. Recent studies
(eg. \citealt{coil}; \citealt{polo}) also show evidence for luminosity
dependence at higher redshifts ($z \simeq 1$).

The results for the colour and morphology dependence of the average
filamentarity are somewhat different from those for the luminosity
dependence. We find that the red galaxies and the ellipticals have a
higher filamentarity as compared to the blue galaxies and spirals
respectively at low filling factors (Figure \ref{fig:9}). There is a
cross-over at $FF \sim 0.15$ for the colour dependence and $FF \sim
0.25$ for the morphology dependence. The blue galaxies and the spirals
have a higher filamentarity as compared to the red galaxies and the
ellipticals respectively at large filling factors. Considering the
length and the thickness (Figure \ref{fig:9}), we find that the length
shows a statistically significant colour and morphology dependence
only at large $FF$ (beyond percolation) where the distribution of blue
galaxies and the spirals has a greater length as compared to red
galaxies and ellipticals respectively. The red galaxies and
ellipticals have a thicker distribution, the differences being
statistically significant at all $FF$. Considering the voids, we find
that the distribution of red galaxies has fewer voids which are larger
(diameter $\sim 127 \pm 46 \, h^{-1} \, {\rm Mpc}$) whereas the
distribution of blue galaxies has more voids which are smaller
(diameter $\sim 98 \pm 17 \, h^{-1}\, {\rm Mpc}$). The ellipticals
and spirals show similar differences with void diameters $\sim 115 \pm
50 \, h^{-1}\, {\rm Mpc}$ and $\sim 90 \pm 21 \, h^{-1}\, {\rm Mpc}$
respectively. These estimates of the void diameter are quite reliable
as the sample size (bin 4) is larger than the void diameter (Figure
\ref{fig:3} and \ref{fig:4}).

We note that our estimates are somewhat larger compared to
results using other methods. For example, \citet{hag} have found that
the voids have a scale of around  $40 \, h^{-1}\, {\rm Mpc}$ in
IRAS 1.2 Jy redshift survey.  A similar conclusion is reached by 
\citet{ul} from their analysis of the Northern Local Void. A recent
analysis of voids in the 2dFGRS by \citet{hoyle} find that voids have
typical diameters of $\sim 30 \, h^{-1}\, {\rm Mpc}$. 
Most of these  estimates  use the area or volume to determine  the
void diameter whereas we use the void perimeter. Substructures in  
the void perimeter is possibly the dominant reason why we get a larger
estimate of the diameter. Further, the large variance in the void
diameter seems to indicate that they have a large spread in sizes. 

The colour and morphology are very strongly correlated galaxy
properties, the red galaxies being predominantly ellipticals and the
blue ones spirals. It is well known that elliptical are found
primarily in dense groups and clusters \citep{dress} whereas the
spirals are distributed in the field. Our finding that at large
length-scales the structures traced by the ellipticals are thicker,
and less filamentary as compared to the spirals is consistent with a
picture where the entire galaxy population is distributed along an
interconnected network of filaments encircling voids, the Cosmic Web.
The ellipticals preferentially inhabit the dense groups and clusters
which can be identified with the nodes of the Cosmic Web, the places
where filaments intersect. The spirals, on the other hand, are
sparsely distributed along the filaments. The colour and morphology
are also known to be correlated with the luminosity. The more luminous
galaxies are predominantly ellitpicals and the fainter ones spirals.

\begin{figure}
\rotatebox{0}{\scalebox{.9}{\includegraphics{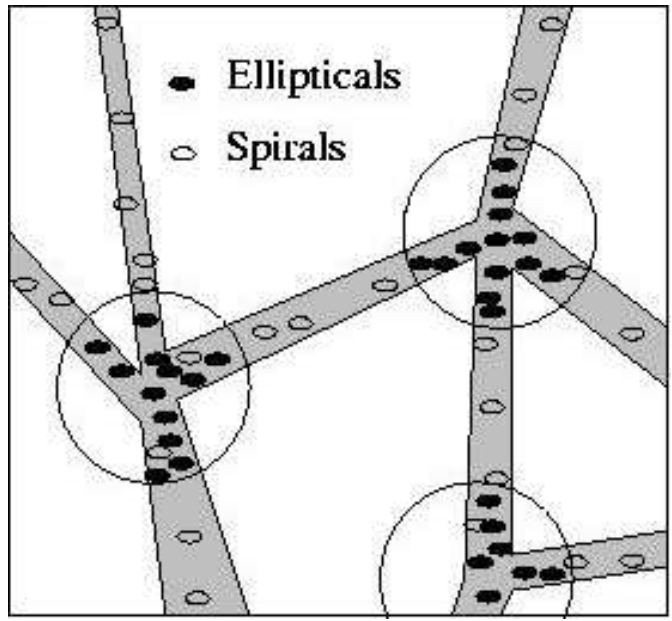}}}
\caption{This shows our picture for the relative distribution of
  ellipticals and spirals along the Cosmic Web. The ellipticals
  densely   inhabit the nodes and are distributed within  the
  encircled regions  shown in the figure. The spirals are sparsely
  distributed along the larger volume of the entire filaments. 
}
\label{fig:10}
\end{figure}

The fact that at small filling factors (which we may associate with
 small length-scales) the eliipticals have a more filamentary
 distribution as compared to spirals, whereas the opposite is found at
 large length-scales is quite intriguing. We discuss below a possible
 interpretation of this finding.  The higher filamentarity of the
 ellipticals at low $FF$ seem to indicate that in addition to residing
 in the clusters at the intersection of filaments, the ellipticals
 also extend a little along the filaments which originate from the
 clusters.  The spirals on the other hand, are distributed along the
 entire extent of the filaments.  This is shown schematically in
 Figure \ref{fig:10}.  In this picture the ellipticals have a more
 compact distribution, and as a consequence they connect up to form
 filamentary clusters at the initial stages of coarse graining. Since
 these structures are localized near the nodes of the filaments and
 they do not extend along the entire length of the filaments, their
 filamentarity increases slowly during the later stages of coarse
 graining. The spirals, on the other hand, are sparsely distributed
 along the entire length of the filaments. They connect up only at a
 later stage of coarse graining to define the entire filamentary
 network of the Cosmic Web.  This is a possible explanation why at low
 filling factors the ellipticals have a higher filamentarity than the
 spirals. Further, it also indicates that with increasing filling
 factor (later stages of coarse graining) the filamentarity of the
 spirals should grow faster than that of the ellipticals as seen in
 (Figure \ref{fig:7}).  Finally we note that the qualitative picture
 presented here is a plausible model which is consistent with the
 quantitative findings of this paper. We do not claim that it is the
 only possible explanation, and it is presented here more in the
 spirit of a hypothesis rather than a conclusion.  Further work is
 required to establish or refute this picture.  

As already mentioned several times, it should be noted, 
that the curves showing $F2$ as a function of $FF$ are not
 absolute. They depend on the geometry of the volume (both shape and
 size) and the galaxy number density. It is only meaningful to compare 
different  galaxy samples for which these  quantities are the
same. Once these are fixed, we can attribute changes in the filamentarity 
to  factors like luminosity, colour or morphology which we are
 testing for.

 Observational findings indicate a close connection between the local
density and galaxy type \citep{dress} with an increase in the  elliptical
and S0 population and a  decrease in the spirals in galaxy clusters.
Our findings are consistent with this, and it indicates that there is
an excess of ellipticals on scales larger than the 
clusters extending to some extent into the filaments. In the
Zel'dovich pancake 
scenario filaments form at the intersection of sheets and clusters
form at the nodes of the filaments. Structure formation occurs in a
hierarchical fashion with sheets forming first, matter flows along the
sheets into the filaments which in turn drain into the clusters. The higher
galaxy density and a denser hot gas environment are possibly
the factors  leading to a higher elliptical fraction in the vicinity of
the nodes.

The relation between our findings and the various theories for galaxy
formation is an important issue. We have compared our findings against
the Millennium Run \citep{springel} which is one of the largest
simulated galaxy catalogues that  incorporates the theory of galaxy
formation in a semi-analytic fashion. The semi-analytic model has a
number of parameters which have been tuned to match various observed
properties of the local galaxy distribution like the luminosity
functions and the Tully-Fisher relation. In this paper we have
compared the filamentarity of the galaxy distribution in the
semi-analytic model against the actual data from the SDSS DR4. 
Studying this in different luminosity bins we find that the two are 
consistent for  $L^{*}$ galaxies, while the brighter galaxies  show
significant differences. For the brighter galaxies,  at large filling
factors   the semi-analytic 
model predicts a substantially higher filamentarity as compared to the
actual data, whereas it underpredicts the same at small $FF$. Though 
the semi-analytic model correctly predicts
the number density of very luminous galaxies  as reflected in the
luminosity function, it fails to correctly reproduce the geometry of
their spatial  distribution. For these galaxies the model does not
predict a sufficiently filamentary distribution at small length-scales 
whereas there is an excess filamentarity predicted at large scales. 

Comparing the filamentarity predicted by the model against the actual
data for galaxies of different colours , we find that the two are
consistent for  red galaxies whereas the distribution of the blue
galaxies are quite different. For these galaxies the model  markedly
underpredicts the filamentarity at small length-scales or filling
factors. We speculate that the discrepancy may be a consequence of two
possible ingredients of the model, the first being a radio mode
feedback from AGNs in massive halos introduced to stop cooling flows.
This effectively quenches  star formation in the galaxies in  
these halos thereby transforming them into red galaxies. 
The  second possibility  is the prescription (plane parallel slab model)
of \citet{kofman5} to include the effects of dust when calculating the
galaxy luminosities and colours. 
 It may be noted  that  \citet{springel} have  found 
discrepancies in the colour dependence of the two point correlation
function. They find that the differences in the correlation amplitudes
of the red and blue galaxies predicted by the model are in excess of
that seen in the 2dFGRS and SDSS. 

In conclusion we note that the filamentary pattern seen in the galaxy
distribution exhibits statistically significant luminosity, colour and
morphology dependence. We have speculated on a  geometrical
picture which can explain some features of the colour and morphology
dependence of the filamentarity. Explaining the origin of the observed
luminosity, colour and morphology dependence in terms of the theories
of galaxy formation is an important issue which needs to be addressed
in the future.

\section{Acknowledgment}

SB would like to acknowledge financial support from the Govt.  of
India, Department of Science and Technology (SP/S2/K-05/2001). BP
would like to thank the CSIR, Govt. of India for financial support
through a Senior Research Fellowship. BP acknowledges Darren Croton
for his help in analyzing the Millennium catalouge. The authors thank
an anonymous referee for the constructive comments and the
comprehensive and detailed review of the paper.

The SDSS DR4 data was downloaded from the SDSS skyserver
http://skyserver.sdss.org/dr4/en/.

    Funding for the creation and distribution of the SDSS Archive has been 
provided by the Alfred P. Sloan Foundation, the Participating 
Institutions, the National Aeronautics and Space Administration, the 
National Science Foundation, the U.S. Department of Energy, the Japanese 
Monbukagakusho, and the Max Planck Society. The SDSS Web site is 
http://www.sdss.org/.

    The SDSS is managed by the Astrophysical Research Consortium (ARC) for 
the Participating Institutions. The Participating Institutions are The 
University of Chicago, Fermilab, the Institute for Advanced Study, the 
Japan Participation Group, The Johns Hopkins University, the Korean 
Scientist Group, Los Alamos National Laboratory, the Max-Planck-Institute 
for Astronomy (MPIA), the Max-Planck-Institute for Astrophysics (MPA), New 
Mexico State University, University of Pittsburgh, Princeton University, 
the United States Naval Observatory, and the University of Washington.

The Millennium Run simulation used in this paper was carried out by
the Virgo Supercomputing Consortium at the Computing Centre of the
Max-Planck Society in Garching. The semi-analytic galaxy catalogue is
publicly available at http://www.mpa-garching.mpg.de/galform/agnpaper

\end{document}